\pdfoutput=1
\documentclass[sigconf,nonacm]{acmart}
\AtBeginDocument{%
  \providecommand\BibTeX{{%
    \normalfont B\kern-0.5em{\scshape i\kern-0.25em b}\kern-0.8em\TeX}}}

\usepackage{dafny}
\usepackage{caption}
\usepackage{subcaption}
\usepackage{fancyvrb}
\usepackage{xcolor}
\usepackage{alltt}
\usepackage{balance}

\begin{document}

\title{Leveraging Large Language Models to Boost Dafny's Developers Productivity}


\author{\'Alvaro Silva}
\affiliation{%
  \institution{Independent Researcher}
  \city{Porto}
  \country{Portugal}
}
\email{amfpsilva@hotmail.com}

\author{Alexandra Mendes}
\affiliation{%
 \institution{HASLab / INESC TEC \& Faculty of Engineering, University of Porto}
  \city{Porto}
  \country{Portugal}
}
\email{alexandra@archimendes.com}

\author{Jo\~{a}o F. Ferreira}
\affiliation{%
  \institution{INESC-ID \& IST, University of Lisbon}
  \city{Lisbon}
  \country{Portugal}
}
\email{joao@joaoff.com}


\begin{abstract}
This research idea paper proposes leveraging Large Language Models (LLMs) to enhance the productivity of Dafny developers. 
Although the use of verification-aware languages, such as Dafny, has increased considerably in the last decade, these are still not widely adopted. Often the cost of using such languages is too high, due to the level of expertise required from the developers and challenges that they often face when trying to prove a program correct. Even though Dafny automates a lot of the verification process, sometimes there are steps that are too complex for Dafny to perform on its own. One such case is that of missing lemmas, i.e. Dafny is unable to prove a result without being given further help in the form of a theorem that can assist it in the proof of the step. 

In this paper, we describe preliminary work on a new Dafny plugin that leverages LLMs to assist developers by generating suggestions for relevant lemmas that Dafny is unable to discover and use. Moreover, for the lemmas that cannot be proved automatically, the plugin also attempts to provide accompanying calculational proofs.
We also discuss ideas for future work by describing a research agenda 
on using LLMs to 
increase the adoption of verification-aware languages in general, by increasing developers productivity and by reducing the level of expertise required for crafting formal specifications and proving program properties.

\end{abstract}





\maketitle

\section{Introduction}
As software is becoming increasingly pervasive in our daily lives, it is more important than ever to ensure that it does not contain bugs. Particularly in critical systems, software testing is not enough as stronger assurances are needed. These assurances can be achieved through software verification, which mathematically ensures that software behaves exactly as intended, with respect to a formal specification. Several verification-aware programming languages and systems, where logical constructs such as pre and postconditions, invariants, and assertions provide assurances about the correctness of the program, are available and enable verification alongside code development. One such language is Dafny \cite{leino2010dafny}.

Dafny is celebrated for its advanced deductive verification support that offers a sophisticated backend, encompassing a compiler capable of producing executable binaries and a verifier tasked with meticulously validating code conformity to specified requirements. 
 While Dafny stands as a state-of-the-art tool in software verification, its application demands a profound grasp of concepts often encountered exclusively during the formalization of specifications, which is not commonplace in conventional programming languages. This complexity can hinder software development productivity in Dafny. 
In addition, and although Dafny has considerable adoption in industry (e.g. at Amazon Web Services\footnote{https://github.com/aws/aws-encryption-sdk-dafny} and at Consensys~\cite{cassez2021verification,cassez2022deductive,cassez2023formal}), it is not widely used, most likely due to the cost in effort, time, and, as stated above, the need for expert knowledge when using Dafny.  

Large Language Models (LLMs) have recently demonstrated extraordinary capabilities, exhibiting proficiency in diverse tasks such as engaging in conversations, retrieving and summarizing extensive information, and even generating and explaining text and code \cite{ai2023gpt}. Their utility is further underscored by their application as code suggestion tools, exemplified by GitHub Copilot \cite{nguyen2022empirical}. 

In this research idea paper, we describe how we plan to explore LLMs' capabilities to boost Dafny's developers productivity and to support its wider adoption. First, we describe some of the challenges identified in the use of Dafny and 
how we intend
to address them. Second, we showcase preliminary work on prompting GPT-4~\cite{ai2023gpt,bubeck2023sparks} to support the automated inference of lemmas and their proofs. Finally, we discuss implications of these results, challenges in using these tools within the context of Dafny, and the next steps of our research agenda. Our proposed solution encompasses the integration of the proposed features into the Dafny plugin for VS Code, seamlessly integrating the presented ideas to assist developers in using Dafny to write specifications and their corresponding implementations. 

Our overarching goal is to enhance the accessibility of Dafny for new users and cultivate increased autonomy and productivity in Dafny's use. 

\section{Synergies between Dafny and LLMs}



\vskip 1em
LLMs and Dafny verification capabilities can complement each other, as LLMs are inherently creative but not always reliable, whilst Dafny's verification engine acts as a stringent gatekeeper, ensuring that artifacts generated by LLMs undergo a rigorous verification process. This interplay makes their combination promising. 

There are several challenges that Dafny users face when using the language. An important challenge is the common case that non-trivial Dafny developments entail substantial effort in writing lemmas and proofs. For example, in Cassez's verification of the Incremental Merkle Tree Algorithm~\cite{cassez2021verification}, almost 90\% of the lines of code are proofs and function definitions used in the proofs. 
Moreover, an experience report on using Dafny
at the VerifyThis 2021 verification competition shows that the interpretability of Dafny's error messages is challenging as these are often not informative~\cite{farrell2021using}.





In this section, we describe four challenges that we intend to tackle using LLMs.

\subsection{Predicate and Lemma Inference}\label{subsec:LemmaInference}

Even though Dafny automates a lot of the verification process, sometimes there are steps that are too complex for Dafny to perform on its own. One such case is that of missing lemmas, i.e. Dafny is unable to prove a result without being given further help in the form of a theorem that can assist it in the proof of the step. 
The challenge lies not only in proving the lemmas and theorems but also in determining the specifications of lemmas and predicates capable of solving the problem at hand. 

Various approaches have been explored in the literature to address these challenges across a diverse range of verification tools. For instance, infering inductive invariants in TLA+~\cite{schultz2022plain}, inference of lemmas in Coq~\cite{sarracinometaprogramming,sivaraman2022data} and in symbolic-Heap
separation logic~\cite{ta2017automated}.

Our objective is to address lemma inference and general predicate inference using 
LLMs. 
Our approach involves using prompting to infer lemmas and predicates, and fine-tuning LLMs to infer them, particularly if the zero-shot or few-shot learning approaches yield unsatisfactory results.
Section \ref{sec:Experiments} provides an illustrative example of lemma inference using prompting and GPT-4~\cite{ai2023gpt}.

\subsection{Proof Inference}
Dafny supports calculational proofs (aka verified calculations), which are proofs by stepwise formula manipulation.
As pointed out by Leino and Polikarpova~\cite{leino2013verified}, calculational proofs are
praised for their rigor, readability, and elegance. Indeed, it has been shown that calculational proofs can greatly improve on traditional verbose proofs in natural language~\cite{ferreira2009students,ferreira2009mathematics,ferreira2011logic}.

Writing calculational proofs is challenging and any method that can help infer these proofs or steps of these proofs can boost Dafny's developers productivity. Moreover, proof inference can complement the lemma inference described in \autoref{subsec:LemmaInference}, as when lemmas are added to the code, it is often the case that a proof needs to be provided by the user. LLMs can assist in this process by providing the full proof or by giving hints to the user regarding the proof steps. This also applies to lemmas inferred by the LLM, as proofs will also need to be provided. In addition, even though Dafny can often prove the lemmas on its own, it has been shown that, even in those cases, adding proof steps that are not absolutely necessary can reduce considerably the verification time~\cite{cassez2021verification}.

To infer proofs, we plan to use models fine-tuned on proof data, but we will also explore few-shot prompted and even zero-shot prompted approaches. Section~\ref{sec:Experiments} provides an illustrative example of few-shot prompted proof generation. 

\subsection{Automated Repair}
Most programmers make mistakes when writing code. Automated Program Repair (APR) can help with this by supporting developers with automatic fix of software bugs. There are several existing works that successfully applied LLMs for automated program repair~\cite{APR_LLMs_2023, jin2023inferfix, QuixBugs_LLM2022, AlphaRepair2023}. However, as far as we are aware, there are no previous developments on APR for Dafny leveraging LLMs. 
In addition, in the context of Dafny, when a bug is detected, i.e. the program fails to verify, the issue can be due to an incorrect specification or an incorrect implementation. Although most current research assumes that the specification is correct, focusing on repairing the implementation, it is known that many issues with software stem from incorrect specifications~\cite{Leveson20}. 
Previous work on specification repair in Dafny relies on Daikon~\cite{ernst2007daikon} for dynamic invariant inference which is then used for generating weakening and strengthening
candidate fixes (and their combination)~\cite{DafnySpecRepair}. 

We are not aware of any work on Dafny's specification or program repair using LLMs. Our goal is to explore existing approaches that use LLMs for automated repair and adapt/improve them to be used for Dafny. 
We also intend to tackle proof repair and to do so in at least two contexts: when the code changes and existing proofs become invalid and when the user or the LLM suggest a proof that does not verify. We will follow a continuous feedback loop between LLMs and the Dafny's verifier, where the LLM produces proofs or fixes to proofs and, should these not verify, the feedback produced by Dafny is fed into the LLM to produce another suggestion.


\subsection{Summarization and Natural Language Specs}
Code summarization, the task of generating summaries that accurately describe the functionalities of the code, can reduce developers' efforts in interpreting the goals of a program or snippet of code. In the context of Dafny, code summarization can also potentially help with understanding the intentions set out in specifications. Further, as detailed above, Dafny's error messages can be challenging to understand; code summarization has the potential to enhance error messages with further information about the error and about what is the mismatch between the specifications and the code. 
As developers spend around 58\% percent of their time on program comprehension activities~\cite{ProgramComprehension2018}, features that assist them in code comprehension seem valuable to enhance their productivity. 

We intend to explore LLMs' capabilities in code summarization to enhance error messages in Dafny, to provide explanations of specs/code, and also complement and update specs/code comments which are helpful in code understanding activities. Previous works have shown the enormous potential of LLMs in code summarization tasks~\cite{CodeDocsGeneration2023, BertCodeSummarization2020, TransformerCodeSummarization2023}.
Another feature that we intend to explore is the use of LLMs to support the translation of natural language specifications into Dafny, assisting developers in this task that requires knowledge that is not commonplace.

\begin{figure*}
\flushleft
\begin{minipage}[t]{\columnwidth}
\begin{subfigure}[t]{\columnwidth}
\begin{Verbatim}[commandchars=\\\@\!]
method CoincidenceCount(a: array<int>, 
                        b: array<int>) 
 returns (c: nat)
 requires forall i,j :: 
   0<=i<j<a.Length ==> a[i]<=a[j]
 requires forall i,j :: 
   0<=i<j<b.Length ==> b[i]<=b[j]
 ensures c == |multiset(a[..]) * multiset(b[..])| {
 c := 0;
 var m, n := 0, 0;
 while m < a.Length && 
       n < b.Length
 invariant 0 <= m <= a.Length && 0 <= n <= b.Length
 @\bfseries\color@red!@// Fails to prove the following invariant!!
 invariant c + |multiset(a[m..]) * multiset(b[n..])|
           == |multiset(a[..]) * multiset(b[..])|
 decreases a.Length - m + b.Length - n {
  if {
   case a[m] == b[n] => {
    \color@blue!@/* Suggested by GPT-4: */!
    \color@blue!@LemmaIntersectionAfterIncrease_mn(a, b, m, n);!
    c, m, n := c + 1, m + 1, n + 1;}
   case a[m] < b[n] => { 
    \color@blue!@/* Suggested by GPT-4: */!
    \color@blue!@LemmaIntersectionAfterIncrease_m(a, b, m, n);!
    m := m + 1;}
   case b[n] < a[m] => {
    \color@blue!@/* Suggested by GPT-4: */!
    \color@blue!@LemmaIntersectionAfterIncrease_m(a, b, m, n);!
    n := n + 1; }}}}
\end{Verbatim}
\caption{Method that computes how many elements two sorted arrays have in common. The second invariant cannot be verified automatically. GPT-4 extends the code with calls to lemmas that enable the  verification.}
\label{fig:CoincidenceCountCode}
\end{subfigure}
\end{minipage}
\hfill
\begin{minipage}[t]{\columnwidth}
\begin{subfigure}[t]{\columnwidth}
\tt
You are a software expert specializing in formal methods using the Dafny programming language. You receive the following program where a loop invariant could not be proven. The verifier error message is inside // VERIFIER ERROR ... //. Your task is to create lemmas and insert them into the code to facilitate verification.
\caption{Prompt used to generate lemmas.}
\label{fig:prompt-lemmas}
\end{subfigure}
\vskip 1em
\begin{subfigure}[b]{\columnwidth}
  \centering
\begin{Verbatim}[commandchars=\\\@\!]
lemma LemmaIntersectionAfterIncrease_m
  (a: array<int>, b: array<int>, m: nat, n: nat)
requires 0 <= m < a.Length && 0 <= n <= b.Length
requires forall i,j :: 0<=i<j<a.Length ==> a[i]<=a[j]
requires forall i,j :: 0<=i<j<b.Length ==> b[i]<=b[j]
ensures |multiset(a[m..]) * multiset(b[n..])| == 
        |multiset(a[m+1..]) * multiset(b[n..])|
        
lemma LemmaIntersectionAfterIncrease_n
  (a: array<int>, b: array<int>, m: nat, n: nat)
requires 0 <= m <= a.Length && 0 <= n < b.Length
requires forall i,j :: 0<=i<j<a.Length ==> a[i]<=a[j]
requires forall i,j :: 0<=i<j<b.Length ==> b[i]<=b[j]
ensures |multiset(a[m..]) * multiset(b[n..])| == 
        |multiset(a[m..]) * multiset(b[n+1..])|

lemma LemmaIntersectionAfterIncrease_mn
  (a: array<int>, b: array<int>, m: nat, n: nat)
requires 0 <= m < a.Length && 0 <= n < b.Length
requires forall i,j :: 0<=i<j<a.Length ==> a[i]<=a[j]
requires forall i,j :: 0<=i<j<b.Length ==> b[i]<=b[j]
\color@gray!@//ensures |multiset(a[m..]) * multiset(b[n..])| ==! 
\color@gray!@          |multiset(a[m+1..]) * multiset(b[n+1..])|!
\color@blue!@ensures |multiset(a[m..]) * multiset(b[n..])| ==! 
\color@blue!@        |multiset(a[m+1..]) * multiset(b[n+1..])|\bfseries@ + 1!!
\end{Verbatim}
%
 \caption{Lemmas generated by GPT-4. The commented assertion in gray had to be replaced with the assertion in blue.}
  \label{fig:lemmas-generated}
\end{subfigure}
\end{minipage}
\caption{GPT-4 suggests lemmas to complete method verification.}
\label{fig:lemma-inference}
\end{figure*}
\section{Preliminary Experiments}\label{sec:Experiments}
This section describes preliminary experiments on using LLMs to infer lemmas and calculational proofs. In particular we used the latest version of GPT-4 Turbo,
gpt-4-1106-preview, trained on data up to April 2023, enabling support for contexts with 128,000 tokens.

\subsection{Lemma Inference}
We use an example taken from Leino's book~\cite{leino2023program} and shown in \autoref{fig:CoincidenceCountCode}. Given two sorted arrays, the method \verb@CoincidenceCount@ computes how many elements they have in common. The postcondition is expressed in terms of multisets (note that \inlinedafny{*} denotes multiset intersection and the vertical-bar brackets denote the cardinality of a multiset). Since the two arrays are sorted, the algorithm can be efficiently implemented using two indices (\inlinedafny{m} and \inlinedafny{n}) to keep track of how many elements of \inlinedafny{a} and \inlinedafny{b} have been processed.
Dafny's verifier is able to prove the method postcondition from the loop invariants. However, it cannot automatically prove the second invariant (identified by the red comment), because it is not able to automatically prove relevant properties about multisets. The typical approach to solve this  is to annotate the program with lemmas that provide enough information for the proof to be completed. 

To determine whether GPT-4 can assist us by inferring useful lemmas, we used the code shown in \autoref{fig:CoincidenceCountCode} 
and
the prompt shown in \autoref{fig:prompt-lemmas}\footnote{The code used was the one shown in black, with the additional comment above the invariant that cannot be proved: \texttt{// VERIFIER\_ERROR loop invariant violation. This invariant could not be proved to be maintained by the loop //}.}. GPT-4 inferred the lemmas shown in \autoref{fig:lemmas-generated} and placed them in the code as shown in \autoref{fig:CoincidenceCountCode} (in blue).
Only a small correction was required to make the program verify using these lemmas as axioms (highlighted in blue). 
However, note that these lemmas cannot be proved since they require more information in the precondition
(e.g. to prove \verb@LemmaIntersectionAfterIncrease_mn@, it is required to add \verb@a[m] == b[n]@ as a precondition; similar annotations are needed for the other lemmas).

\subsection{Proof Inference}
Our experiments with prompt engineering to infer proofs for the three lemmas shown in \autoref{fig:lemmas-generated} were not as successful. In general, there were many syntactic errors and we had to provide many examples of proofs to generate plausible solutions. The best result we obtained was when we asked to prove one of the lemmas, but gave a complete proof of one of the other similar lemmas.

To test whether GPT-4 would be able to help Dafny developers when they need to prove less domain-specific results, we also attempted to infer proofs for statements that are more widely known.
An example is shown in \autoref{fig:linear-combination-proof}, where a proof for the lemma \inlinedafny{Factor0} was generated by GPT-4. The lemma is a version of a basic property of number theory: $p$ is a factor of any linear combination $p*a + p*b$.
The predicate is a variation of the lemmas required to prove correctness of Euclid's algorithm as implemented in Dafny's integration tests.\footnote{Permalink: \url{https://github.com/dafny-lang/dafny/blob/eae8fc97aabc13ea665b486060e831188628d42b/Source/IntegrationTests/TestFiles/LitTests/LitTest/dafny4/gcd.dfy}. See also: \url{https://leino.science/papers/krml279.html}}

The best attempt by GPT-4 to prove the lemma is shown in blue in \autoref{fig:linear-combination-proof}. To achieve this, we designed a prompt incorporating extensive Dafny code and examples; in particular we provided all the contents of the integration test file for calculations.\footnote{Permalink: \url{https://github.com/dafny-lang/dafny/blob/eae8fc97aabc13ea665b486060e831188628d42b/Source/IntegrationTests/TestFiles/LitTests/LitTest/dafny2/Calculations.dfy}}
Even though the hints are correct and make sense, they are not accepted by Dafny. Nevertheless, once we comment them, the proof is accepted. Moreover, in the version of Dafny that we used, 4.3.0, we had to change the variable definitions to:
\[
\verb@var a :| x == p*a; var b :| y == p*b;@
\]


\begin{figure}[t]
\begin{Verbatim}[commandchars=\\\@\!]
type pos = x | 1 <= x witness 1

ghost predicate IsFactor(p: pos, x: pos) {
  exists q :: p * q == x }

ghost function Factors(x: pos): set<pos> {
  set p: pos | p <= x && IsFactor(p, x) }

lemma Factor0(p: pos, y: pos, x: pos)
    requires exists a :: x == p*a
    requires exists b :: y == p*b
    ensures IsFactor(p, y + x) {  
   \color@blue!@/* Proof Suggested by GPT-4: */!
   \color@blue!@ var a: pos := x / p; // Witness for x == p*a!
   \color@blue!@ var b: pos := y / p; // Witness for y == p*b!

   \color@blue!@  calc {!
   \color@blue!@      p * (a + b);!
   \color@blue!@      == { arithmetic }!
   \color@blue!@      p * a + p * b;!
   \color@blue!@      == { definition of a and b }!
   \color@blue!@      x + y; }!}
\end{Verbatim}
\caption{GPT-4-generated proof for lemma \inlinedafny{Factor0}.}
\label{fig:linear-combination-proof}
\end{figure}

\section{Related Work}
Clover~\cite{sun2023clover} is the only work that we are aware of on using LLMs in the context of Dafny. Clover consists on using a checker that performs consistency
checks among code, docstrings, and formal annotations. The main idea is to reduce correctness checking to a problem of consistency checking. However, the reduction from correctness to consistency is not mathematically complete. 
Our goals are different and more fine-grained: 
we aim to prove correctness and to contribute with features that continuously focus on parts of the program, not necessarily the program as a whole, to enhance developers productivity in achieving program correctness.


Regarding lemma inference, we are not aware of any work focused on Dafny nor work that uses LLMs, but there have been efforts to synthesize lemmas for Coq~\cite{sarracinometaprogramming,sivaraman2022data}, and for symbolic-Heap separation logic~\cite{ta2017automated}.
{\sc AdtInd} 
automates proofs by induction over algebraic data types, where lemmas
are synthesized by term enumeration guided by user-specified templates~\cite{yang2019lemma}.

Recent work on neural methods to automate proof synthesis is related to our goals of inferring calculational proofs.
Given a partial proof and the proof state, neural theorem provers use neural networks to predict the next likely proof steps, which are then evaluated by a proof assistant to return new proof states or errors. 
Several neural theorem provers have been proposed for Coq~\cite{sanchez2020generating, thakur2023languageagent,tactician,yang2019learning, First20oopsla, Sanchez-Stern22passport, First22icse,Huang2019Gamepad}, for 
Lean~\cite{yang2023leandojo} and Isabelle~\cite{Jiang2022Thor, mikula2023magnushammer}.
%
Baldur uses LLMs to repair proofs as part of its proof synthesis approach~\cite{first2023baldur}. 
Finally, regarding computer support for calculational proofs, there have been efforts to create structure editors that also support verified calculations~\cite{mendes2014structure,mendes2018towards} but none using LLMs.

\section{Conclusion}
The lemma inference experiment yielded promising results, demonstrating GPT-4's ability to infer lemmas with minor errors. However, throughout the experiments, several responses contained syntax errors that proved challenging to rectify through reprompting. The program struggled to complete lemma proofs or correct minor errors in lemma specifications. 

On the other hand, the proof inference experiment revealed that only through a meticulously crafted prompt with rich examples could GPT-4 successfully complete the proofs. 
This underscores the need for 
improved LLMs that reduce the need for extensive manual prompt curation. 

 In order to achieve these improvements, our next steps will focus on prompting engineering and fine-tuning models with well-crafted datasets containing relevant Dafny code, with a specific emphasis on lemma inference, proof inference, and suggestions that only contain correct syntax. 

In terms of dataset creation, we plan to contribute to and extend CloverBench~\cite{sun2023clover}, the dataset used by Clover. At the time of writing, CloverBench contains only 60 small CS textbook examples. Many more, and more complex, examples will be required to ensure the practicality of our ideas. Moreover, we plan to create additional and specialized datasets for specific tasks; for example, we intend to create a Dafny dataset akin to Reichel et al.'s dataset for Coq~\cite{reichel2023proof}, to 
assist developers in proof repair.

Finally, to increase the adoption of our ideas, we started their integration into Dafny's VSCode plugin. 

\newpage
\bibliographystyle{ACM-Reference-Format}
\bibliography{main,coqpyt}

\end{document}